\documentstyle[12pt]{article}

\def\noi{\noindent}

\def\del{\partial}
\def\nab{\nabla}

\def\dis{\displaystyle}

\begin{document}

\begin{center}
{\Large\bf Third quantization of $f(R)$-type
gravity}
\\[10mm]
\end{center}

\noi
\hspace*{1cm}\begin{minipage}{14.5cm}
Yoshiaki Ohkuwa$^1$ and Yasuo Ezawa$^2$\\

\noi
$^1$ Section of Mathematical Science, Dept. of Social Medicine, Faculty 
of Medicine,\\
$\ $  University of Miyazaki, Miyazaki, 889-1692, Japan\\
$^2$ Dept. of Physics, Ehime university, Matsuyama, 790-8577, Japan\\

\noi
Email : ohkuwa@med.miyazaki-u.ac.jp, ezawa@phys.sci.ehime-u.ac.jp 
\\

\noi
{\bf Abstract}\\[2mm]
We examine the third quantization of $f(R)$-type gravity, 
based on its effective Lagrangian 
in the case of a flat Friedmann-Lemaitre-Robertson-Walker metric. 
Starting from the effective Lagrangian, we execute 
a suitable change of variable and the second quantization, and 
we obtain the Wheeler-DeWitt equation. The third quantization 
of this theory is considered. And the uncertainty relation of the universe 
is investigated in the example of $f(R)$-type gravity, where $f(R)=R^2$. 
It is shown, when the time is late 
namely the scale factor of the universe is large, the spacetime does not  
contradict to become classical, and, when the time is early namely the scale 
factor of the universe is small, the quantum effects are dominating.

\end{minipage}

\section{Introduction}

Since the discovery of the accelerated expansion of the universe
\cite{accel,SN,WMAP}, much attention has been attracted to the generalized gravity 
theories of the $f(R)$-type\cite{CDTT,SF,NO}.
Before the discovery, such theories have been interested in because of its 
theoretical advantages: 
The theory of the graviton is renormalizable.\cite{UDW, Stelle} 
It seems to be possible to avoid the initial singularity of the universe 
predicted by the theorem by Hawking.\cite{Hawking}\cite{NT} 
And inflationary model without inflaton field is possible.\cite{Staro}

On the other hand quantum cosmology is a quantum theory of the universe 
as a whole, and this is described by the Wheeler-DeWitt equation, 
which is a differential equation on the wave function of the universe.\cite{WDW} 
The quantum cosmology of the $f(R)$-type gravity has been already studied.\cite{QCf(R)}  
However, it is well known that, in general, 
the Wheeler-DeWitt equation has the problem 
that the probabilistic interpretation is difficult as in the case of 
the Klein-Gordon equation.  
One of the proposed ideas to solve this problem is the third quantization 
in analogy with the quantum field theory.\cite{3Q} 
Then the third-quantized universe theory describes a system of many universes.
Third quantization is useful to describe bifurcating universes and merging
universes,  
if an interacting term is introduced in the Lagrangian for the 
third quantization.

In this work we examine the third quantization of the $f(R)$-type gravity.
We start from the effective theory of the $f(R)$-type gravity in a flat 
Friedmann-Lemaitre-Robertson-Walker metric. 
Then a suitable change of variable is performed and the Wheeler-DeWitt 
equation is written down. 
Quantizing this model once more, we obtain the third-quantized theory 
of $f(R)$-type gravity. 
The Heisenberg uncertainty relation is investigated in a feasible model 
of the $f(R)$-type gravity, where $f(R)=R^2$. 
It will be shown, when the time is late 
namely the scale factor of the universe is large, the spacetime does not  
contradict to become classical, and, when the time is early namely the scale 
factor of the universe is small, the quantum effects are dominating.

In section 2, the effective theory of $f(R)$-type gravity in the case of
 a flat Friedmann-Lemaitre-Robertson-Walker metric is summarized .
In section 3, the third quantization of this theory is considered.
In section 4, the uncertainty relation is studied 
in the example of $f(R)$-type gravity, where $f(R)=R^2$.
Summary is given in section 5.

\section{Generalized gravity of $f(R)$-type}

Generalized gravity of $f(R)$-type is one of the higher curvature 
gravity in which the action is given by
$$
S=\int d^4 x{\cal L}=\int d^4 x\sqrt{-g}f(R).                      \eqno(2.1)
$$
The spacetime is taken to be $4$-dimensional. 
Here $g\equiv \det g_{\mu\nu}$ and $R$ 
is the scalar curvature.
Field equations are derived by the variational principle as follows
$$
f'(R)R_{\mu\nu}-{1\over2}f(R)g_{\mu\nu}
-\nab_{\mu}\nab_{\nu}f'(R)+g_{\mu\nu}\Box f'(R)=0 ,                \eqno(2.2)
$$
where $f'(R)={d f(R) \over d R}$, 
$\nab_{\mu}$ is the covariant derivative with respect to $g_{\mu\nu}$ 
and $\Box=g^{\mu\nu} \nab_{\mu} \nab_{\nu}$.

Let us consider the next action
$$
S=\int d^4 x \sqrt{-g}\Bigl[f(\phi)+f'(\phi)(R-\phi)\Bigr]  ,      \eqno(2.3)
$$
where $f''(\phi) \neq 0$. 
It is well known that the field equations of this action are Eq.(2.2) 
in which $f(R)$ is replaced with $f(\phi)$ and the following equation 
$$
R=\phi .                                                           \eqno(2.4)
$$
If we substitute Eq.(2.4) into Eq.(2.3), we formally obtain Eq.(2.1).

In order to make things simple, let us consider the case of a flat 
Friedmann-Lemaitre-Robertson-Walker metric,\cite{SFf(R)}
$$
ds^2=-dt^2+ a^2 (t) \sum_{k=1}^3 (dx^k)^2          .               \eqno(2.5)
$$
Then the scalar curvature is written as 
$$
R= 6 \left({\ddot{a} \over a} + {\dot{a}^2 \over a^2} \right) ,    \eqno(2.6)
$$
with $\dot{a}={da \over dt}$.
Substituting Eqs.(2.5),(2.6) into Eq.(2.2), we obtain 
$$
H^2  =  {1 \over 3f'(R)}\left[ {1 \over 2}\Bigl( Rf'(R)-f(R) \Bigr)
-3H \dot{R} f''(R) \right]   ,  \hspace{3.8cm}                     \eqno(2.7) 
$$
$$    
2\dot{H}+3H^2  =  -{1 \over f'(R)} \left[
f'''(R)\dot{R}^2+2H\dot{R}f''(R)+{\ddot R}f''(R) 
+{1 \over 2} \Bigl( f(R)-Rf'(R) \Bigr)
\right] ,                                                          \eqno(2.8)
$$
where $H={\dot{a} \over a}$ .

Using Eqs.(2.5),(2.6), we can write Eq.(2.3) as 
$$
S_{eff}=\int d^4 x \Bigl[ a^3 f(\phi)-6f''(\phi)\dot{\phi}a^2 \dot{a}
-6f'(\phi) a {\dot a}^2 -f'(\phi)\phi a^3 \Bigr]
=\int d^4 x {\cal L}_{\it eff}                                ,    \eqno(2.9)
$$
where the partial integration has been applied. 
The classical equations of motion can be derived from the following equations 
$$
{\del {\cal L}_{\it eff} \over \del a}
-{d \over dt}{\del {\cal L}_{\it eff} \over \del {\dot a}} =0   ,  \eqno(2.10)
$$
$$
{\del {\cal L}_{\it eff} \over \del \phi}
-{d \over dt}{\del {\cal L}_{\it eff} \over \del {\dot \phi}} =0.  \eqno(2.11)
$$
Eqs.(2.6),(2.11) give
$$
\phi=6 \left({{\ddot a} \over a}+{{\dot a}^2 \over a^2} \right) 
= R.                                                               \eqno(2.12) 
$$
We can obtain Eq. (2.8) from Eqs.(2.10),(2.12).
The canonical momenta for $a$ and $\phi$ are defined as
$$
p_a= {\del {\cal L}_{\it eff} \over {\dot a}}
=-12a{\dot a}f'(\phi)-6a^2f''(\phi){\dot \phi}
=-6a^2(2Hf'(\phi)+f''(\phi){\dot \phi})                         ,  \eqno(2.13)
$$
$$
p_{\phi}= {\del {\cal L}_{\it eff} \over {\dot \phi}}
=-6a^2 {\dot a} f''(\phi) = -6a^3Hf''(\phi)  .   \hspace{4.3cm}    \eqno(2.14)
$$
So the Hamiltonian is written as
$$
\begin{array}{ll}
{\cal H}_{\it eff} &\equiv p_a {\dot a} 
+p_{\phi} {\dot \phi} -{\cal L}_{\it eff} ,
 \\[3mm]
&=\dis -{p_a p_{\phi} \over 6a^2 f''(\phi)}
+{f'(\phi)p_{\phi}^2 \over   6a^3f''(\phi)^2}
+a^3 f'(\phi) \phi - a^3 f(\phi)          .        
\end{array}                                                        \eqno(2.15)
$$
The time reparametrization invariance means 
$$
{\cal H}_{\it eff} = 0 .                                           \eqno(2.16)
$$                                                                
This equation gives Eq. (2.7).
Therefore ${\cal L}_{\it eff}$ can be regarded as the effective Lagrangian 
for Eq. (2.1) when the metric is given by Eq. (2.5).

\section{Third quantization}

Now let us make the change of variable as follows:
$$
\varphi=\varphi(\phi)=\ln f'(\phi), \qquad f'(\phi)=e^{\varphi}, \qquad 
\phi=f'^{-1}(e^{\varphi})   .                                      \eqno(3.1)
$$
Then the Lagrangian and canonical momenta become
$$
{\cal L}_{\it eff}=-6a {\dot a}^2e^{\varphi}
-6a^2 {\dot a}e^{\varphi}{\dot \varphi}
-a^3f'^{-1}(e^{\varphi}) e^{\varphi}
 +a^3 f\Bigl( f'^{-1}(e^{\varphi}) \Bigr)  ,                       \eqno(3.2)
$$
$$
p_a={\del {\cal L}_{\it eff} \over \del {\dot a}}
=-12a {\dot a} e^{\varphi}-6a^2 e^{\varphi} {\dot \varphi}   , 
 \hspace{4cm}                                                      \eqno(3.3)
$$
$$
p_{\varphi}={\del {\cal L}_{\it eff} \over \del {\dot \varphi}}
=-6a^2 {\dot a} e^{\varphi}                      .  \hspace{6cm}   \eqno(3.4)
$$
So the Hamiltonian can be written as
$$
\begin{array}{ll}
{\cal H}_{\it eff} &= p_a {\dot a} +p_{\varphi} {\dot \varphi} 
-{\cal L}_{\it eff}, 
\\[3mm]
&=\dis 
{p_{\varphi}^2 \over 6a^3 e^{\varphi}}
-{p_a p_{\varphi} \over 6a^2 e^{\varphi}}
+a^3 f'^{-1}(e^{\varphi}) e^{\varphi}
-a^3  f\Bigl( f'^{-1}(e^{\varphi}) \Bigr)         .        
\end{array}                                                        \eqno(3.5)
$$
If we multiply $6a^2 e^{\varphi}$ to Eq. (3.5), 
the Hamiltonian constraint becomes 
$$
{p_{\varphi}^2 \over a}-p_a p_{\varphi}
+6 a^5 f'^{-1}(e^{\varphi}) e^{2 \varphi}
-6a^5 e^{\varphi}  f\Bigl( f'^{-1}(e^{\varphi}) \Bigr) 
=0                                                      .          \eqno(3.6)
$$
Substituting
$$
p_a \rightarrow -i {\del \over \del a} ,\qquad
p_{\varphi} \rightarrow -i {\del \over \del \varphi} ,             \eqno(3.7)
$$
we obtain the Wheeler-DeWitt equation
$$
-{1 \over a}{\del^2 \psi \over \del \varphi^2}
+{\del^2 \psi \over \del a \del \varphi}
+6a^5 f'^{-1}(e^{\varphi}) e^{2 \varphi} \psi
-6a^5 e^{\varphi} f\Bigl( f'^{-1}(e^{\varphi}) \Bigr) \psi=0.      \eqno(3.8)
$$
Here $\psi$ is the wave function of the universe.

The action for the third quantization 
to yield the Wheeler-DeWitt equation (3.8) is
$$
\begin{array}{ll}
S_{3Q}&=\dis \int d a d \varphi {1 \over 2} \left[
{1 \over a} \Biggl( {\del \psi \over \del \varphi} \Biggr)^2
-{\del \psi \over \del a}{\del \psi \over \del \varphi}
+6a^5 f'^{-1}(e^{\varphi}) e^{2 \varphi} \psi^2 
-6a^5 e^{\varphi} f\Bigl( f'^{-1}(e^{\varphi}) \Bigr) \psi^2
\right]   , \\[3mm]
&=\int d a d \varphi {\cal L}_{3Q}  . 
\end{array}                                                        \eqno(3.9)
$$ 
If we consider $a$ to be the time coordinate, 
the canonical momentum which is conjugate to $\psi$ is written as
$$
p_{\psi}={\del {\cal L}_{3Q} \over \del {\del \psi \over \del a}}
=-{1 \over 2}{\del \psi \over \del \varphi} .                      \eqno(3.10)
$$ 
The Hamiltonian is given by
$$
\begin{array}{ll}
{\cal H}_{3Q}&=\dis p_{\psi}{\del \psi \over \del a}-{\cal L}_{3Q},  \\[3mm]
&=-{2 \over a}p_{\psi}^2-3a^5 f'^{-1}(e^{\varphi}) e^{2 \varphi} \psi^2
+3a^5 e^{\varphi} f\Bigl( f'^{-1}(e^{\varphi}) \Bigr) \psi^2.
\end{array}                                                        \eqno(3.11)                       
$$

In order to third quantize this system we impose the equal time 
commutation relation 
$$
\bigl[ {\hat \psi}(a,\varphi) , {\hat p_{\psi}}(a,\varphi') \bigr] 
= i \delta (\varphi - \varphi')  .                                 \eqno(3.12)
$$
We use the $\rm Schr\ddot{o}dinger$ picture, so we take the operator 
${\hat \psi} (a,\varphi)$ as the time independent c-number field 
$\psi (\phi)$, and we substitute the momentum operator as 
$$
{\hat p_{\psi}} (a,\varphi) \rightarrow 
-i{\del \over \del \psi (\varphi)} .                               \eqno(3.13)
$$
Then we obtain the $\rm Schr\ddot{o}dinger$ equation
$$
i{\del \Psi \over \del a} ={\hat {\cal H}}_{3Q} \Psi   ,           \eqno(3.14)
$$
that is
$$
i{\del \Psi \over \del a} 
=\left[ {2 \over a} \Biggl( {\del \over \del \psi (\varphi)} \Biggr)^2
-3a^5 f'^{-1}(e^{\varphi}) e^{2 \varphi} \psi^2 (\varphi) 
+3a^5 e^{\varphi} f\Bigl( f'^{-1}(e^{\varphi}) \Bigr) \psi^2 (\varphi) 
\right] \Psi ,                                                     \eqno(3.15)
$$
where $\Psi$ is the third quantized wave function of universes.

\section{Uncertainty relation}

We assume the Gaussian ansatz for the solution to the 
$\rm Schr\ddot{o}dinger$ equation (3.14)
$$
\Psi(a,\varphi,\psi (\varphi) )
=C \exp \left\{ -{1 \over 2}A(a,\varphi)[\psi (\varphi)-\eta (a,\varphi)]^2
+iB(a,\varphi)[\psi (\varphi)-\eta (a,\varphi)] \right\}  ,        \eqno(4.1)
$$
where $A(a,\varphi)=D(a,\varphi)+iI(a,\varphi)$.
The real functions 
$D(a,\varphi), I(a,\varphi), B(a,\varphi)$ and $\eta (a,\varphi)$ 
should be determined from Eq. (3.15).
$C$ is the normalization of the wave function.
The inner product of two functions $\Psi_1$ and $\Psi_2$ is defined as 
$$
\langle \Psi_1 , \Psi_2 \rangle 
=\int d \psi (\varphi) \Psi_1^*(a,\varphi,\psi (\varphi))
 \Psi_2(a,\varphi,\psi (\varphi))                                  \eqno(4.2)
$$

Let us calculate Heisenberg's uncertainty relation. 
The dispersion of $\psi (\varphi)$ is defined as
$(\Delta \psi (\varphi))^2 \equiv \langle \psi^2 (\varphi) \rangle
-\langle \psi (\varphi) \rangle^2$.
From Eqs. (4.1),(4.2) we have
$$
\langle \psi^2 (\varphi) \rangle = {1 \over 2D(a,\varphi)} 
+ \eta^2 (a, \varphi) , 
\qquad \langle \psi (\varphi) \rangle = \eta (a,\varphi) ,         \eqno(4.3)
$$
then
$$
(\Delta \psi (\varphi))^2 = {1 \over 2D(a,\varphi)} .              \eqno(4.4)
$$
Similarly the dispersion of $p_{\psi}(\varphi)$ is defined as
$(\Delta p_{\psi}(\varphi))^2 \equiv \langle p_{\psi}^2 (\varphi) \rangle
-\langle p_{\psi}(\varphi) \rangle^2$, so we obtain
$$
\langle p_{\psi}^2 (\varphi) \rangle 
= {D(a,\varphi) \over 2}+{I^2 (a,\varphi) \over 2D(a,\varphi)}
+B^2 (a,\varphi) , 
\qquad \langle p_{\psi} (\varphi) \rangle = B(a,\varphi) ,         \eqno(4.5)
$$
then we have
$$
(\Delta p_{\psi}(\varphi))^2 
= {D(a,\varphi) \over 2}+{I^2 (a,\varphi) \over 2D(a,\varphi)} .   \eqno(4.6)
$$
Therefore the uncertainty relation can be written as
$$
(\Delta \psi (\varphi))^2 (\Delta p_{\psi}(\varphi))^2
={1 \over 4} \Biggl( 1+ {I^2 (a,\varphi) \over D^2 (a,\varphi)} 
\Biggr)   .                                                        \eqno(4.7)
$$

Note that to evaluate (4.7), only $A(a,\varphi)$ 
is necessary. However, to solve the general equation for
$A(a,\varphi)$ is difficult, so let us take a feasible example such as 
$$
f(R) = R^2  .                                                      \eqno(4.8)
$$
In this case 
$$
f'(R) = 2R =e^{\varphi} , \qquad 
f'^{-1}(e^{\varphi}) = R = {e^{\varphi} \over 2} , \qquad
f\Bigl( f'^{-1}(e^{\varphi}) \Bigr) = {e^{2 \varphi} \over 4} .    \eqno(4.9)
$$
The $\rm Schr\ddot{o}dinger$ equation (3.15) becomes
$$
i {\del \Psi \over \del a} 
= \left[ {2 \over a} \Biggl( {\del \over \del \psi (\varphi)} \Biggr)^2
-{3 \over 4}a^5 e^{3 \varphi} \psi^2 (\varphi) \right] \Psi .      \eqno(4.10)
$$
Substituting the ansatz (4.1) into Eq. (4.10), we obtain
$$
-{i \over 2} {\del A(a,\varphi) \over \del a}
={2 \over a} A^2 (a,\varphi) -{3 \over 4} a^5 e^{3 \varphi}  .     \eqno(4.11)
$$
Writing
$$
\ln a = {\alpha \over 6}  ,                                        \eqno(4.12)
$$
we have
$$
-3i{\del A(\alpha,\varphi) \over \del \alpha} 
=2 A^2 (\alpha,\varphi) - {3 \over 4} e^{\alpha} e^{3 \varphi} .   \eqno(4.13)
$$
In order to solve this equation, let us write
$$
A(\alpha,\varphi)={3i \over 2}{\del \ln u(\alpha,\varphi) 
\over \del \alpha} ,                                               \eqno(4.14)
$$
where $u(\alpha,\varphi)$ is a suitable function.
Then $u(\alpha,\varphi)$ must satisfy the equation, 
$$
{\del^2 u(\alpha,\varphi) \over \del \alpha^2}
+{1 \over 6} e^{3 \varphi} e^{\alpha} u(\alpha,\varphi) = 0 .      \eqno(4.15)
$$
Now we write
$$
z=2\sqrt{{e^{3 \varphi} e^{\alpha} \over 6}} ,                     \eqno(4.16)
$$
and we have 
$$
{\del^2 u(z,\varphi) \over \del z^2}
+{1 \over z}{\del u(z,\varphi) \over \del z} 
+ u(z,\varphi) = 0 .                                               \eqno(4.17)
$$
As this equation can be regarded as the ordinary 
differential equation with respect to $z$ 
under the assumption that $\varphi$ is fixed other than in $z$, 
this is the case when $\nu=0$ in the following Bessel's equation
$$
{d^2 u(z) \over d z^2}
+{1 \over z}{d u(z) \over d z} 
+\left( 1- {\nu^2 \over z^2} \right) u(z) = 0 .                    \eqno(4.18)
$$ 
Therefore we have the solution
$$
u(z,\varphi)=c_J(\varphi) J_0 (z) + c_Y(\varphi) Y_0 (z)      ,    \eqno(4.19)                        
$$
where $J_0, \ Y_0$ are the Bessel functions of order $0$ 
and $c_J, \ c_Y$ are arbitrary complex functions of $\varphi$. 

From Eqs. (4.9),(4.12),(4.14),(4.16),(4.19), we can obtain 
$z={4 \over \sqrt{3}} R^{3 \over 2} a^3,\ \varphi=\ln (2R)$ and 
$$
A(z,\varphi)=-i {3z \over 4} 
{c_J(\varphi) J_1 (z) + c_Y(\varphi) Y_1 (z) 
\over c_J(\varphi) J_0 (z) + c_Y(\varphi) Y_0 (z) },    
                                                                   \eqno(4.20)
$$
where we have used $J'_0(z)=-J_1(z), \ Y'_0(z)=-Y_1(z)$\cite{AS}.

Since $A(z,\varphi)=D(z,\varphi)+iI(z,\varphi)$ , we can derive
$$
D(z,\varphi)= -{3i \over 4\pi 
\vert c_J(\varphi) J_0 (z) + c_Y(\varphi) Y_0 (z) \vert^2}
[c_J(\varphi) c_Y^*(\varphi) -c_J^*(\varphi) c_Y(\varphi)]  ,      \eqno(4.21)
$$
where we have used 
$J_0(z)Y_1(z)-J_1(z)Y_0(z)=-{2 \over \pi z}$\cite{AS}, and 
$$
\begin{array}{ll}
I(z,\varphi)=\dis &-{3z \over 8  
\vert c_J(\varphi) J_0 (z) + c_Y(\varphi) Y_0 (z) \vert^2}\times \\[3mm]
&\Bigl[ 2\vert c_J(\varphi) \vert^2 J_0(z) J_1(z)
+2\vert c_Y(\varphi) \vert^2 Y_0(z)Y_1(z)  \\[3mm]
&+(c_J(\varphi) c_Y^*(\varphi) + c_J^*(\varphi) c_Y(\varphi))
(J_0(z) Y_1(z) + J_1(z) Y_0(z)) \Bigr]  .                          
\end{array}                                                       \eqno(4.22)
$$

So if we assume 
$c_J(\varphi) c_Y^*(\varphi) -c_J^*(\varphi) c_Y(\varphi) \neq 0$ 
(Note that in this case both of $c_J(\varphi), c_Y(\varphi)$ are nonzero.), 
we can obtain 
$$
\begin{array}{ll}
\dis {I^2 (z,\varphi) \over D^2 (z,\varphi)}= 
&-{\pi^2 z^2 \over 
4[c_J(\varphi) c_Y^*(\varphi) -c_J^*(\varphi) c_Y(\varphi)]^2}\times \\[3mm]
&\Big[ 2\vert c_J(\varphi) \vert^2 J_0(z) J_1(z)
+2\vert c_Y(\varphi) \vert^2 Y_0(z)Y_1(z)  \\[3mm]
&+(c_J(\varphi) c_Y^*(\varphi) + c_J^*(\varphi) c_Y(\varphi))
(J_0(z) Y_1(z) + J_1(z) Y_0(z)) \Bigr]^2  .                          
\end{array}                                                       \eqno(4.23)
$$

When the time is late namely $a \rightarrow \infty$
i.e. $z \rightarrow \infty$,
$$
\begin{array}{ll}
\dis J_0(z) &\sim \sqrt{{2 \over \pi z}} \cos (z-{\pi \over 4}) ,
\qquad J_1(z) \sim \sqrt{{2 \over \pi z}} \sin (z-{\pi \over 4}) , \\[3mm]
Y_0(z) &\sim \sqrt{{2 \over \pi z}} \sin (z-{\pi \over 4}) ,
\qquad Y_1(z) \sim -\sqrt{{2 \over \pi z}} \cos (z-{\pi \over 4}) ,
\end{array}                                                     
$$
\cite{AS}
we have
$$
\begin{array}{ll}
{I^2 (z,\varphi) \over D^2 (z,\varphi)} \dis &\sim 
-{\left[( \vert c_J(\varphi) \vert^2-\vert c_Y(\varphi) \vert^2) \cos (2z)
+(c_J(\varphi) c_Y^*(\varphi) +c_J^*(\varphi) c_Y(\varphi)) \sin (2z) \right]^2 
\over [c_J(\varphi) c_Y^*(\varphi) -c_J^*(\varphi) c_Y(\varphi)]^2} \\[3mm]
&\sim O(1) . 
\end{array}                                                      \eqno(4.24)
$$
This and Eq. (4.7) mean that when time is late that is  
in the course of the universe expansion  
namely $a \rightarrow \infty$, 
the spacetime does not contradict to become classical.

On the other hand when the time is early namely
$a \rightarrow 0$ i.e. $z \rightarrow 0$,
$$
\begin{array}{ll}
\dis J_0(z) &\sim 1-{z^2 \over 4} ,
\qquad J_1(z) \sim {z \over 2} ,  \\[3mm]
Y_0(z) &\sim {2 \over \pi}\Bigl( \ln {z \over 2}+\gamma \Bigr) , 
\qquad Y_1(z) \sim -{2 \over \pi z} , 
\end{array}                          
$$
\cite{AS}
we obtain
$$
{I^2 (z,\varphi) \over D^2 (z,\varphi)} \sim 
-{16 \vert c_Y(\varphi) \vert^4 \over 
\pi^2 [c_J(\varphi) c_Y^*(\varphi) -c_J^*(\varphi) c_Y(\varphi)]^2}
\Bigl( \ln {z \over 2} +\gamma \Bigr)^2 \sim \infty  ,           \eqno(4.25)
$$
where $\gamma$ is a constant.
This and Eq. (4.7) mean that the fluctuation of the third quantized universe 
field becomes large when the time is early namely 
$a \rightarrow 0$. 
Therefore the quantum effects are dominating for the small values of the 
universe radius.

\section{Summary}

In this work the third quantization of the $f(R)$-type gravity is investigated, 
when the metric is a flat Friedmann-Lemaitre-Robertson-Walker metric.
The Heisenberg uncertainty relation is investigated in a feasible model 
of the $f(R)$-type gravity, that is $f(R)=R^2$ . 
It has been shown in this model, when the time is late 
namely the scale factor of the universe is large, the spacetime does not  
contradict to become classical, and, when the time is early namely the scale 
factor of the universe is small, the quantum effects are dominating. 
As a future work, it will be interesting to investigate a more realistic model 
such as $f(R)=R+c R^2$, where c is a constant.

\section*{Acknowledgements}

One of the authors (Y.O.) would like to thank Prof. H. Ohtsuka 
for letting know him a clue to solve Eq. (4.15).

\section*{Note added}

After this work has been done, we have been notified 
the recent papers on the third quantization 
\cite{Faizal1, Faizal2, Calcagni} by their authors. 
In Ref. \cite{Faizal1} the third quantization of f(R) gravity 
was also investigated.

\end{document}